\documentclass[10pt,a4paper,twocolumn,showkeys]{revtex4-2}

\usepackage{mathptm}
\usepackage{graphicx} 
\usepackage{amsmath,amssymb} 
\usepackage{color}
\begin{document}
	
\title{Quantum droplets in dipolar  quasi-one-dimensional Bose-Einstein condensates in optical lattices}
\author{Sk Siddik}\email{amanandsafin@gmail.com}
\author{Golam Ali Sekh}\email{skgolamali@gmail.com}
\affiliation{Department of Physics, Kazi Nazrul University, Asansol-713340, W.B., India}  
\begin{abstract} 
We consider quantum droplets in  dipolar Bose-Einstein condensates embedded in optical lattices within the framework of  Gross-Pitaevskii equations. In dipolar BECs, the long-range and isotropic dipole-dipole interaction provides an additional mechanism for self-binding. We analyze the linear stability  as well dynamics of quantum droplets.  We find an effective potential for the width and show that the optimum width for formation of quantum droplet increases as the dipole-dipole interaction increases. We  study dynamics of the stable droplets and see that its width oscillates with time,  and the amplitude of oscillation increases with the increase of dipole-dipole interaction. In presence of optical lattices, width of a stable droplet changes quasi-periodically while  the location of density profile changes periodically in space. The frequency of oscillation  depends sensitively on the lattice parameters. We check dynamical stability of a quantum droplet  through numerical simulation.
	\end{abstract} 
\maketitle   
\section{Introduction}  
Quantum droplets (QDs) are considered as the manifestation of  beyond  mean-field effects in quantum many-body systems. Unlike ordinary liquid droplets, the QDs are very dilute and remain stable due to  balance between attractive mean-field interaction and repulsive beyond mean-field effects known as the Lee–Huang–Yang (LHY) correction \cite{a1, a2}. Quantum droplets have been originally predicted in Bose-Bose mixtures near the collapse regime by Petrov et al \cite{a1} and realized  experimentally in homo-nuclear potassium condensates \cite{a6, a7} and  hetero-nuclear  potassium–rubidium (K–Rb)\cite{a8} and sodium–rubidium (Na–Rb) condensates mixtures \cite{a9}.
	
In dipolar Bose-Einstein condensate, the long range dipole-dipole interactions (DDIs), quantum fluctuations and mean-field interaction interplay to generate quantum droplets (QDs). 
Particularly, the QDs are generated  in the strongly dipolar regime of the condensate where QDs are stablized due to  quantum fluctuation instead of collapse \cite{pra1, pra2, pra3}. 
The QDs have been observed experimentally  in  the ultra-cold dipolar gases of dysprosium and erbium atoms, displaying pronounce anisotropy and self-bound behaviors which are distinct from their binary counterparts \cite{a3, a4, a5}.  The presence of long-range  DDI in a BEC with LHY corrections led to several studies  including vortex QDs \cite{nr1,nr2,nr3} and dipolar supersolid \cite{nr4,nr5,nr6}.

An optical lattice (OL) is a periodic potential created through interference of counter propagating laser beams. Studies on BECs with artificial periodic potential  has been receiving  a great interest due  to the experimental flexibility  to control lattice parameters \cite{bb17}.  A dipolar BEC with OLs can support different phases such as   supersolid, Mott insulator and collapse phases  through modifications of experimentally accessible parameters \cite{aa11, aa12,nr7}. In the recent past, the stability of standard dipole-model QDs in optical lattices \cite{nr8} and multi-stability of QDs in OLs have been investigated \cite{nr9}, and shown that the OLs help stabilize the QDs. Recently, quench and Floquet dynamics of quantum droplets have been studies in two-dimensional optical lattices \cite{nr9a}.

Our objective in this paper is to study  the dynamics of quantum droplets (QDs) in a dipolar Bose-Einstein condensates with optical lattice. More specifically, we consider quasi-one-dimensional BECs and find the effects of DDI on the effective potential of QDs. The  width corresponding to the potential minimum i.e., the optimum width  for the formation of a QD  increases with the increase of DDI strength.  In presence of  optical {lattices}, the minimum value of effective potential of width is augmented keeping the optimum width almost same. We show that the   width of a QD oscillates quasi-periodically in time  and the amplitude of oscillation increases with the increase of DDI strength.  The density profile  oscillates  in space about the lattice minimum and  shape of  the distribution depends sensitively on the lattice parameters. These droplets are found to satisfy Vakhitov-Kolokolov criterion  \cite{nr10,nr11,nr12}. They can be linearly  and dynamically stable for chosen values of parameters.

The paper is organized as follows. In section II, we introduce a model for quantum droplet in dipolar quasi-one-dimensional BECs  loaded optical lattices and find equations for different parameters of a flat-top QD using variational approach. In section III, we  discuss  the properties of self-bound droplets without an optical lattice potential.   We investigate, in Section IV, the effects of optical lattices on the properties of the self-bound droplets and study dynamics of QDs by  numerical simulation. Finally, we summarize our results and give some remarks in section V.

\section{Theoretical formulation} 
The quantum fluctuation in Bose-Einstein  condensate of dipolar  is the reason for the generation of quantum droplets. Theoretically, the effects of quantum fluctuation is incorporated  in the Gross-Pitaevskii (GP) equation through Lee-Huang-Yang (LHY) correction. The modified GP equation in quasi-one-dimensional geometry describing  dipolar Bose-Einstein condensates is given by \cite{f1,f2,f3,f4,f4a,f4b,f34} 
\begin{equation}
i\hbar\frac{\partial \psi}{\partial t}\!= \!\left(\! -\frac{\hbar^{2}}{2m} \frac{\partial^{2}}{\partial x^{2}} +V_{\rm {ext}}\left( x\right) - g |\psi|^{2}+g_{1}|\psi|^{3}+\phi_{\rm{dd}}\right) \psi. 
\label{Eq1a}
\end{equation} 
Here $\psi(x,t)$ is the order parameter, and $g={2 \hbar^2 a_{s}}/( {m l_\perp^2})$, $g_{1}={256\hbar^2 a^{5/2}}(1+\frac{3}{2} \epsilon_{\rm dd}^2)/({15 \pi m l_\perp^3})$, $V_{\rm ext}$ and $\phi_{\rm{dd}}$ are the strength of mean-field effect, quantum fluctuation, external potential and dipole-dipole interaction respectively.  Understandably, $l_{\perp}=\sqrt{{\hbar}/{\left(m\omega_{\perp} \right) }}$ is the length of transverse harmonic trap of frequency $\omega_{\perp}$ and  $\epsilon_{\rm dd}=a_{dd}/a_{s}$ with  $a_{s}$ and $a_{dd}$ represent respectively $s$-wave scattering length and  dipole-dipole interaction length.

 The external potential in Eq.(\ref{Eq1a}) is given by 
\begin{equation}
V_{\rm{ext}}\left( x\right) =\frac{1}{2}m\omega^{2} x^{2}+V_{0} \cos^{2}\left(k_{L} x \right).
\end{equation} 
Here $\omega$ is the frequency in the axial direction and $V_{0}$ is the strength of optical lattice with wave number $k_L$. The dipole-dipole interaction is given by
\begin{equation}
\phi_{\rm{dd}}=\int dx' V^{1D}_{\rm{dd}}\left(x-x' \right) \left|\psi\left(x',t \right)  \right| ^{2} .
\label{Eq3q}
\end{equation}  
 Note that the quasi-one-dimensional regime is obtained by taking $\omega<<\omega_{\perp}$ and considering mean-field energy much smaller than transverse confinement energy. Under this condition, the transverse distribution of the condensate is described by $\exp\left[\frac{-\left(y^{2}+z^{2} \right) }{2l_{\perp}} \right]/\left(l_{\perp}\sqrt{\pi} \right)$. Integrating the three-dimensional DDI over the transverse direction using transverse distribution yields the  following effective one-dimensional DDI \cite{f4a,f5,f6,f7}. 
\begin{equation}
V^{1D}_{\rm{dd}}\left(x-x' \right) =U_{\rm{dd}}{V}_{\rm{dd}}\left(|x-x'|\right),
\end{equation}  
where {$U_{\rm{dd}}=\frac{C_{\rm{dd}}\left[1+3\cos(2\theta) \right] }{32 \pi l^{3}_{\perp}}$ with $\,\,\,C_{dd}=\frac{\mu_{0}g^{2}_{L}\mu^{2}_{B}}{4\pi}$, where $\mu_{0}$ is the permeability in free space, $g_{L}$ is the Land\'e factor, $\mu_{B}$ is the Bohr magneton  and 
\begin{eqnarray}
	  	{V}_{\rm {dd}}\left(x \right) &=&\left[2|x|/l_{\perp}-\sqrt{2\pi}\left(1+x^{2}/l^{2}_{\perp} \right){\rm e}^{x^{2}/2l^{2}_{\perp}} {\rm{erfc}}\left(|x|/\sqrt{2}l_{\perp} \right)   \right]\nonumber \\  &+&\frac{8}{3}\delta \left(x/l_{\perp} \right). 
	  \label{e5}
\end{eqnarray}}
Here ${\rm erfc}$ stands for complementary error function and $\theta$ is the angle of polarization. For far-field approximation  $|x-x'|>>l_{\perp}$,  Eq.(\ref{e5}) turns out to be {\cite{36a}}
\begin{equation}
{V}_{\rm{dd}}\approx \frac{2l_{\perp}^3}{|x-x'|^{3}}.
\end{equation} 
{For short-range approximation $x\rightarrow x',$ ${V}_{\rm{dd}}\approx\frac{8l_{\perp}}{3}\delta (x) $. Substituting this in Eq. (3), we see that it gives an effect two-body contact interaction $C_{\rm dip}$ and, therefore, $g$ in Eq. (1) is replaced by $g_{\rm e}=g+C_{\rm dip}$.}

We introduce the following  variables,  $$ \tilde{x}=\frac{x}{l_{\perp}},\,\,\, \tilde{t}=\omega_{\perp} t \,\,\,\,{\rm and}\,\, \,\,\tilde{\psi}=\sqrt{\frac{l_{\perp}}{N}}\psi. $$ 
 and obtain the dimensionless version of Eq. (\ref{Eq1a})
	\begin{equation}
		i\frac{\partial \tilde{\psi}}{\partial t}-\left(  -\frac{1}{2} \frac{\partial^{2}}{\partial x^{2}} +\tilde{V}_{\rm {ext}}\left(\tilde{x} \right)  - {\tilde{g}_{\rm{e}}}|\tilde{\psi}|^{2}+\tilde{g}_{1}|\tilde{\psi}|^{3}+\tilde{\phi}_{\rm{dd}}\right) \tilde{\psi} =0.
		\label{Eq5}
	\end{equation} 
Here $\tilde{V}_{\rm{ext}}\left(\tilde{x} \right)=\frac{1}{2}\rho^{2}\tilde{x}^{2}+\tilde{V}_{0}\cos^{2}\left(\eta\tilde{x} \right) $ and $\tilde{\phi}_{dd}= \tilde{U}_{\rm{dd}}\int d\tilde{x'} \tilde{V}_{\rm{dd}}\left(|\tilde{x}-\tilde{x'}| \right)|\tilde{\psi}\left(\tilde{x'} \right)|^{2}$ with $\tilde{V}_{\rm{dd}}(\tilde{x})=2|\tilde{x}|-\sqrt{2\pi}(1+\tilde{x}^2)\, {\rm exp}(\tilde{x}^2/2)\,{\rm erfc}(|\tilde{x}|/\sqrt{2})$. Here $\rho=\frac{\omega}{\omega_{\perp}}$ is the aspect ratio between the axial ($\omega$) and  transverse ($\omega_{\perp}$) trapping frequencies, $N$ is the number of atoms, and $k_L l_{\perp}=\eta$. The parameters in Eq.(\ref{Eq5}) are given by 
	$$\tilde{V}_{0}=\frac{V_{0}}{\hbar \omega_{\perp}},\, \tilde{g}_{\rm{e}}=\frac{g_{\rm{e}} N}{\hbar \omega_{\perp} l_{\perp}},\,  \tilde{g_{1}}=\frac{g_{1}N^{3/2}}{\hbar \omega_{\perp} l^{3/2}_{\perp}},\,\tilde{U}_{\rm{dd}}=\frac{N U_{\rm{dd}}}{\hbar\omega_{\perp} }.$$
In the rest of the work, we omit tildes from Eq.(\ref{Eq5}) for convenience of presentation.

The Lagrangian density is given by
	\begin{eqnarray}
		\mathcal{L}&=&\frac{i}{2}\left(\psi^{*}\psi_{t}-\psi\psi^{*}_{t} \right) -\frac{1}{2}\left|\psi_{x} \right| ^{2} -V_{\rm{ext}}\left( x\right)|\psi|^{2}+\frac{g}{2}|\psi|^{4} \nonumber \\ &-&\frac{2g_{1}}{5}|\psi|^{5}-\frac{1}{2}\phi_{\rm{dd}}|\psi|^{2}.
		\label{Eq7a}
	\end{eqnarray} 
 We consider the following super Gaussian trial solution to describes a quantum droplet \cite{f8,f3, nr13}.  
\begin{equation}
\psi=A\,\, {\rm e}^{-\frac{1}{2}\left(\frac{\left(x-x_{0} \right) }{w} \right) ^{2\sigma}+i b\left(x-x_{0} \right)^{2}+i k(x-x_{0})+i\phi }. 
\label{Eq7}
\end{equation} 
Here $A(t),\,\, w(t),\,\, b(t),\,\, k(t),\,\,\, \phi(t)$ and $x_0(t)$ are the time-dependent variational parameters, denoting respectively amplitude, width, chirp, wavenumber, phase and center-of-mass of the QD. The norm  given by,  
\begin{equation}
P=\int_{-\infty}^{+\infty} \left|\psi\left(x,t \right)  \right| ^{2} dx =\left|  A \right| ^{2} \frac{w}{\sigma}\Gamma\left( \frac{1}{2\sigma}\right), 
\label{Eq9a}
\end{equation} 
is a conserved quantity and it is proportional to number of atoms in the droplet. In Eq.(\ref{Eq7}), $\sigma$ can be varied to generate various types of quantum droplets. For example, Eq.(\ref{Eq7})  gives sharp-top QD for $\sigma=1$ while it gives flat-top QD  for $\sigma\geq 2$. Understandably, flatness of a droplet increases as $\sigma$ increases. {In this context, we note that the shape of a sharp-top QD coincides with that of a bright soliton but their origins are quite different. Particularly, there is no role of quantum fluctuation in the formation of matter-wave bright solitons \cite{36b}}.  Here, we have chosen to work with flat-top QDs ($\sigma=2$) and calculate $\tilde{\phi}_{\rm{dd}}$, the effective dipole-dipole interaction (DDI). The DDI  is given by  
	\begin{equation}
		\phi_{\rm{dd}}= \frac{C_{0}}{|x-x_{0}|^{3}} \sum^{\infty}_{n=0} \frac{(2n+1)(2n+2)}{2}\frac{w^{2\sigma}}{(x-x_{0})^{2n}} \frac{\Gamma(\frac{2n+1}{2\sigma})}{\Gamma(\frac{1}{2\sigma})}.
	\end{equation} 
Here $C_{0}=\frac{C_{\rm{dd}}\left[1+3\cos(2\theta) \right]N }{16 \pi \,\,\hbar \omega_\perp l_\perp^3}$. The averaged Lagrangian defined by $L=\int_{-\infty}^{+\infty}\mathcal{L}dx$ is calculated using  Eq.(\ref{Eq3q}) and Eq.(\ref{Eq7}). For $\sigma=2$, $L$ is given by  
\begin{equation}
L=-P\dot{\phi}+P k\dot{x_{0}}-\frac{P \Gamma(3/4)}{4\Gamma(5/4)}\left[ w^{2}\dot{b}+2b^{2}w^{2}\right] +G(P, x_{0}, w).
\end{equation} 
Where $$G(P, x_{0}, w)=-\frac{K_{0}}{w^{2}}+\frac{g_{\rm{e}} K_{1}}{w}-\frac{g_{1}K_{2}}{w^{3/2}} +H(P, w, x_{0})-G_{\rm{dd}}(w,P),$$
with  
\begin{equation}
K_{0}=\frac{P\Gamma(7/4)}{2\Gamma(5/4)},\,\, K_{1}=\frac{P^{2}}{2^{5/4}\Gamma(1/4)},\,\,\,	K_{2}=\frac{2^{11/4}}{5^{5/4}} \frac{P^{5/2}}{\left( \Gamma(1/4)\right) ^{3/2}}, \nonumber
\end{equation}  
\begin{eqnarray}
&&H\left(P, w, x_{0} \right) =-\frac{PV_{0}\cos\left(2\eta x_{0} \right)}{2} \,\,  _{0}F_{2}\left(;\frac{1}{2},\frac{3}{4};\frac{\eta^{4}w^{4}}{16} \right)\\ \nonumber &-&\frac{PV_{0}}{2}-\frac{Pk^{2}}{2}+\frac{P V_{0}\eta^{2}\Gamma(\frac{3}{4})w^{2}\cos\left(2\eta x_{0} \right)}{\Gamma(1/4)}\,\,  _{0}F_{2}\left(;\frac{5}{4},\frac{3}{2}; \frac{\eta^{4}w^{4}}{16} \right),\nonumber
\end{eqnarray} 
\begin{equation}
G_{\rm{dd}}=\frac{P C_{0}}{2\sqrt{\pi}\left( \Gamma(1/4)\right) ^{2}}\sum^{\infty}_{n=0} \frac{(2n+1)(2n+2)}{{w^{2n+2}}} \frac{\Gamma(\frac{(2n+1)}{4})}{{\left[\Gamma(-\frac{(2n+1)}{2})\right]^{-1}}}. \nonumber
 \end{equation}
	
The equations for the variational parameters obtained from the vanishing condition of $\frac{\delta L}{\delta y_j}$, for {$y_{j}= \phi,\,\,\, b,\,\,\,k,\,\,\,x_{0}$ and $ k$,}
   can be  written as follows. 
\begin{equation}
\frac{d}{dt}\left(\frac{1}{2}\Gamma\left(1/4 \right)|A|^{2} w  \right)=0,
\end{equation}  
\begin{equation}
\ddot{w}=\frac{2\Gamma(1/4)}{P\Gamma(3/4)} \frac{\partial G}{\partial w}=-\frac{dU}{dw},
\end{equation} 
with  
\begin{eqnarray}
U=\frac{2\Gamma(1/4)}{P \Gamma(3/4)}\left[ \frac{K_{0}}{w^{2}}-\frac{g_{\rm{e}}K_{1}}{w}+\frac{g_{1}K_{2}}{w^{3/2}} -H+G_{\rm{dd}} \right],
\end{eqnarray}
{\begin{eqnarray}
		\dot{k}&=& V_{0}\eta\sin(2\eta x_{0})\,\, _{0}F_{2}\left(; 1/2,3/4;\frac{\eta^{4}w^{4}}{16} \right)\\ \nonumber &-& \frac{2V_{0}\eta^{3}w^{2}\Gamma(3/4)}{\Gamma(1/4)}\sin(2\eta x_{0})\,\, _{0}F_{2}\left(;5/4,3/2;\frac{\eta^{4}w^{4}}{16} \right)
		\label{eq17aa} 
\end{eqnarray} and} 
{\begin{equation}
		\dot{x}_{0}=k
		\label{eq18a}
\end{equation} } 
{In Eq.(17), $_{0}F_{2}$ stands for a special function of the generalized hypergeometric series. Combining Eqs.(17) and (\ref{eq18a}), we obtain }
\begin{eqnarray}
\ddot{x}_{0}=-\frac{dU_{\rm{OL}}}{dx_{0}} 
\end{eqnarray} 
	with   
\begin{eqnarray}
U_{\rm{OL}}\left(x_{0} \right) &=& \frac{V_{0} \cos\left(2\eta x_{0} \right)}{2} \,\,  _{0}F_{2}\left(;\frac{1}{2},\frac{3}{4};\frac{\eta^{4}w^{4}}{16} \right)\\ \nonumber &-&\frac{V_{0} \eta^{2}\Gamma(\frac{3}{4})w^{2}\cos\left(2\eta x_{0} \right)}{\Gamma(1/4)}\,\,  _{0}F_{2}\left(;\frac{5}{4},\frac{3}{2}; \frac{\eta^{4}w^{4}}{16} \right).\nonumber
\end{eqnarray} 
Eq. (14) implies that the number of atom in a QD is not changing with time while its width and center-of-mass vary with time. Particularly, width of the droplet moves in a effective potential given in Eq. (16) and center-of-mass faces an effective potential given in Eq. (20). For a small value of $x_0$, the center-of-mass executes harmonic oscillation and the frequency of oscillation is given by
\begin{eqnarray}
\Omega=2\eta\sqrt{{V_{0}{ \cos(2\eta x_m)}}\left(-\frac{1}{2}\,\, Y_{1} +\frac{\eta^{2} w^{2}\Gamma(3/4)}{\Gamma(1/4)}\,\, Y_{2} \right)}.
\end{eqnarray}  
Here $Y_{1}=\, _0F_{2}\left(;\frac{1}{2},\frac{3}{4};\frac{\eta^{4}w^{4}}{16} \right)$, $Y_{2}=\,_{0}F_{2}\left(;\frac{5}{4},\frac{3}{2};\frac{\eta^{4}w^{4}}{16} \right)$ and  $x_m$ is the position of lattice minimum. 

{In order to check linear stability, we find chemical potential ($\mu$) by calculating energy through $E=\left\langle\left( \frac{i}{2}\left(\psi^{*}\psi_{t}-\psi\psi^{*}_{t} \right)-\mathcal{L}\right) \right\rangle$ for the ansatz in Eq.(\ref{Eq7}) 
 and using $\mu=\frac{E}{P}$. The chemical potential is given by}
\begin{equation}
\mu= \alpha_{0}+\alpha_{1} w^{2}-\frac{\alpha_{2}P}{w}+\frac{\alpha_{3}P^{3/2}}{w^{3/2}}+\frac{\alpha_{4}}{w^{2}}+\alpha_{5} +\mu_{\rm{dd}}. 
	\end{equation}
	Here  
	\begin{equation}
		\alpha_{0}=\frac{V_{0}}{2}+\frac{k^{2}}{2},\,\,\, \alpha_{1}= \frac{2b^{2}\Gamma\left( 3/4\right) }{\Gamma\left(1/4 \right) },\,\,\,\alpha_{2}=\frac{g_{\rm{e}}}{2^{1/4}\Gamma (1/4)},\nonumber
	\end{equation} 
		\begin{equation}
		\alpha_{3}=\frac{g_{1}2^{11/4}}{5^{5/4}\Gamma(1/4)^{3/2}},\,\,\alpha_{4}=\frac{2 \Gamma\left(\frac{7}{4} \right) }{\Gamma\left(\frac{1}{4} \right) }, \,\, \mu_{\rm{dd}}=\frac{G_{\rm{dd}}}{P},\nonumber
	\end{equation}
	\begin{eqnarray}
		\nonumber
		\alpha_{5}&=&\frac{V_{0}\cos(2\eta x_{0})}{2}\,\, _{0}F_{2}\left(;\frac{1}{2},\frac{3}{4};\frac{\eta^{4}w^{4}}{16} \right) \\ \nonumber &-&\frac{\eta^{2}V_{0}\Gamma(3/4)\cos(2\eta x_{0})w^{2}}{\Gamma(1/4)}\,\, _{0}F_{2}\left(;\frac{5}{4},\frac{3}{2};\frac{\eta^{4}w^{4}}{16} \right). \nonumber
	\end{eqnarray}
According to Vakhitov-Kolokolov criterion, a QD is linearly stable if $d\mu/dP$ is negative \cite{nr10,nr11,nr12}.

{In the following sections, we discuss the properties of quantum droplets based on the theoretical formulation. However, it is quite useful to estimate the values of different parameters from some real experiments. In a typical experiment of $^{164}{\rm Dy}$ \cite{a3}, the values of $\omega_{\perp}\sim 2\pi\times 133$ rad/sec, $l_{\perp}\sim 6.8\times 10^{-7}$m,  $a_s\sim 4.23\times 10^{-9}$ m and $\epsilon_{dd}\sim 1.83$. This  gives $\tilde{g}_1 \approx 0.75$ and $\tilde{g}_e \approx 4.75$ for $N\sim400$.} 
	   	 
\section{QDs in dipolar Bose-Einstein condensate} 
The dipole-dipole interaction plays a significant role on the formation and stability of quantum droplet. To illustrate the effects of DDI, we plot in
\begin{figure}
\centering
\includegraphics[scale=0.3]{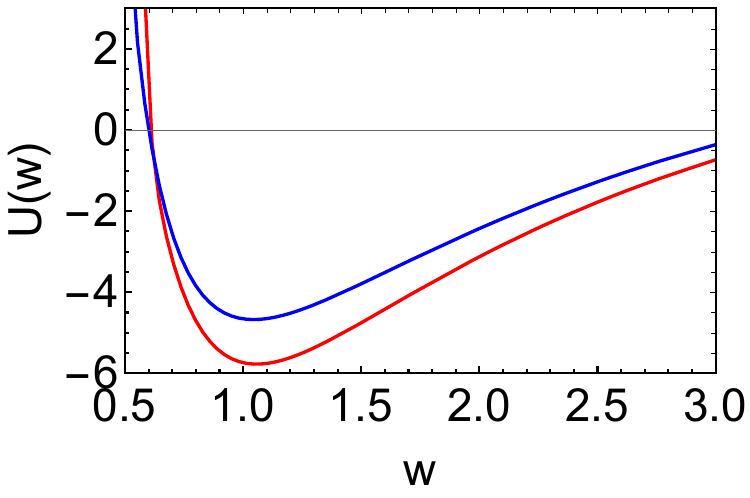} 
\includegraphics[scale=0.28]{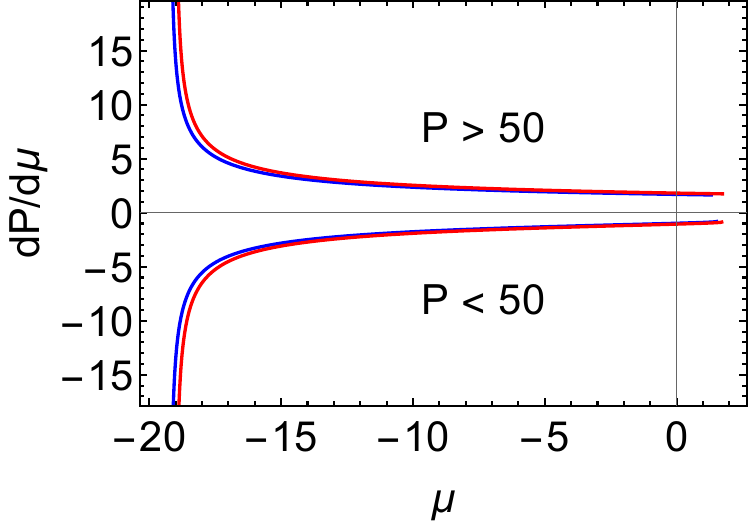} 
\caption{Left panel: Variation of effective potential with width for different values  of dipole-dipole interaction in absence of optical lattice. The blue curve is drawn for $C_{0}=0.01$ and red curve is drawn $C_{0}=1.7$.  Right panel: {Variation of $\frac{dP}{d\mu}$ as a function of chemical potential $\mu$ for $w$ values corresponding to potential minima.} In both the panels, we have taken $P=10,\,\, g_{\rm{e}}=4.75,\,\, g_{1}=0.75,\,\, V_{0}=0$ and $ k=1.5$.} 
\label{fig1}
\end{figure}  
Fig.\ref{fig1}, the effective potential of $U$ with $w$.  We see that as DDI increases, the optimum width for the formation of QD increases and the depth of the potential also increases (left panel). It means that the stronger dipole interaction makes the droplet robust against compression caused by attractive mean-field interaction \cite{pra1,pra2}. {We see that $dP/d\mu$  remains negative as long as $P$ remains below $50$. This indicates that  the droplet is linearly stable. However,  the droplet becomes mechanically unstable as $P\geq 50$.}  With the increase of DDI,  the chemical potential changes and the stability  of a QD can further be influenced.
	 \begin{figure}
	 	\centering
	 	\includegraphics[scale=0.3]{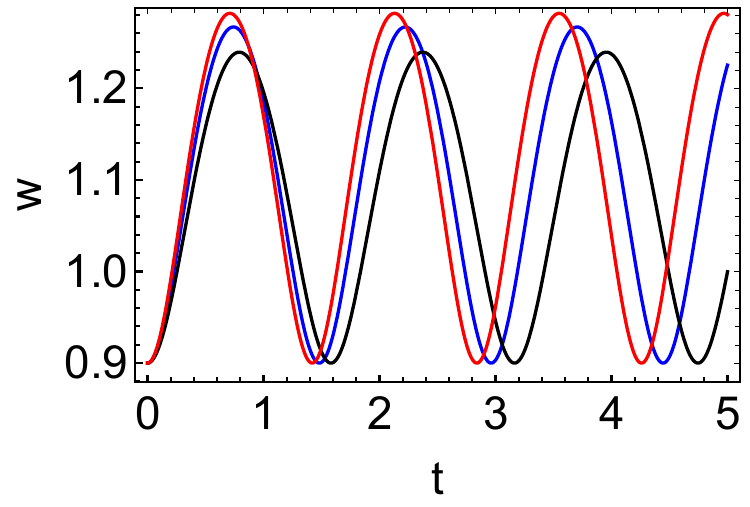} 
	 	\includegraphics[scale=0.34]{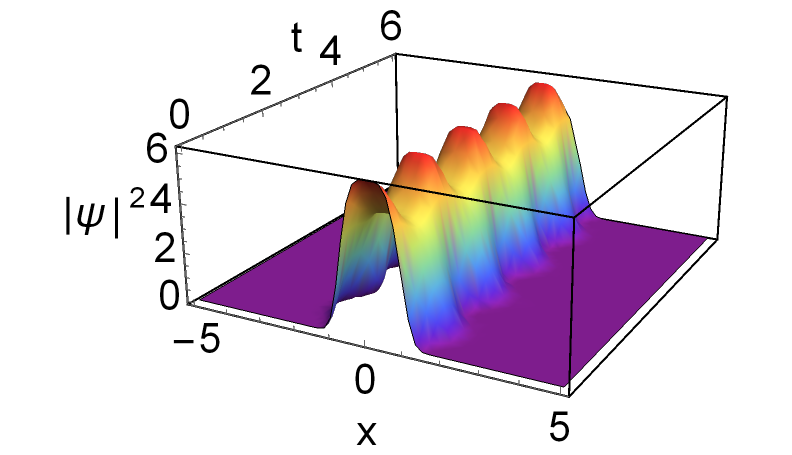} 
	 	\caption{Left panel: Variation of width with time for different values  of dipole-dipole interaction in absence of optical lattice. The red curve is drawn for $C_{0}=1.7$, blue curve is drawn for $C_{0}=1$, and black curve is drawn $C_{0}=0.01$. Right panel: Evolution of density profile for the parameters corresponding to the red curve and $x_0=0.05$.  In both the panel, we have taken $P=10,\,\, g_{\rm{e}}=4.75,\,\, g_{1}=0.75,\,\, V_{0}=0,$ and $k=1.5$.} 
	 	\label{fig2}
	 \end{figure} 
	 \begin{figure}
	 	\centering
	 	\includegraphics[scale=0.27]{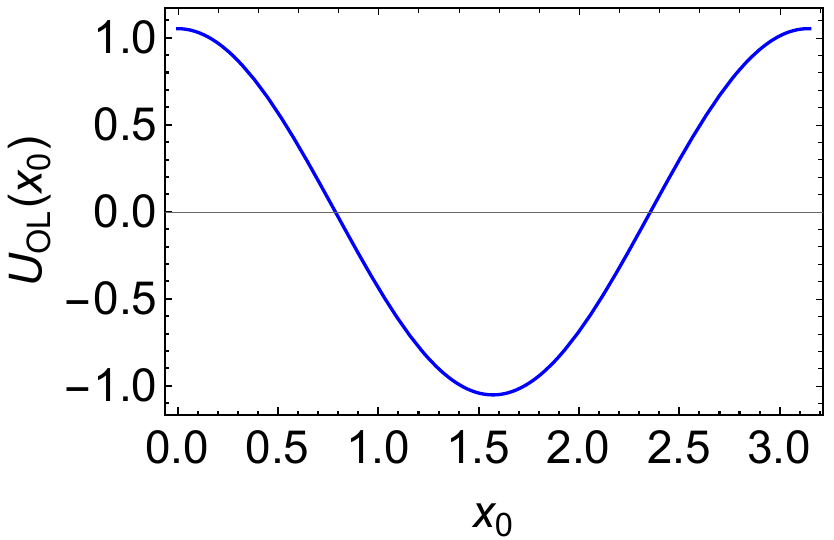} 
	 	\includegraphics[scale=0.28]{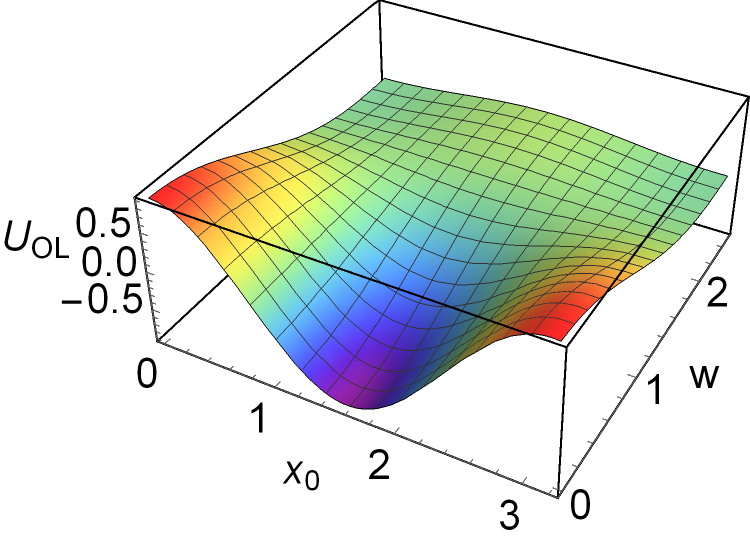} 
	 	\caption{Left panel: Variation of effective optical lattice potential with the change of center of mass for $ w(0)=1.0$. Right panel: Variation of effective optical lattice potential with width and center of mass. In both the panel, we take  $ V_{0}=1.8$ and $\eta=1.0$. } 
	 	\label{fig3}
	 \end{figure}  
	 
In order to find the effects of DDI on the dynamics of QDs, we solve Eq. (15) in absence of optical lattices ($V_0=0$). It is seen that the width of a QD oscillates and the amplitude of oscillation  increases with the increase of dipole-dipole interaction ({Fig.\ref{fig2}}, left panel). The change of width results in the variation of density profile of QDs with time ({Fig.\ref{fig2}}, right panel).
	  
\section{QD in optical lattices} 
The optical lattice can change effective potential and thus affect the properties of quantum droplets in dipolar Bose-Einstein condensates. We see that the occurrence of minimum in effective potential depends on the location of center of mass ($x_0$)  in the lattice and width of the QD. Particularly, {the minimum at $x_0=1.6$ disappears} as the width becomes relatively large (Fig.3).  With a view to find the effect of OL, {we take $x_0=0.5$} and vary the strength of OL. We see that the depth of the potential minimum decreases with increase of lattice strength. However, the optimum width for the formation of QD remains unchanged (left panel, Fig. 4). {The QDs can be linearly stable  if the number of atoms remains below $60$ while the droplet becomes unstable for $P\geq 60$. The stability curves  for different values of $V_0$ differ if $\mu < 0$ and they approach each other if $\mu > 0$ (right panel, Fig. 4)}.  	  
 \begin{figure}
 	\centering 
 	\includegraphics[scale=0.3]{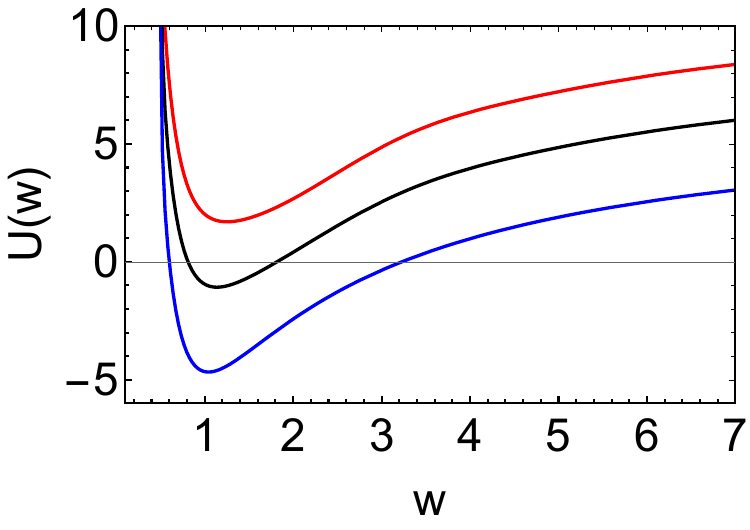} 
 	\includegraphics[scale=0.28]{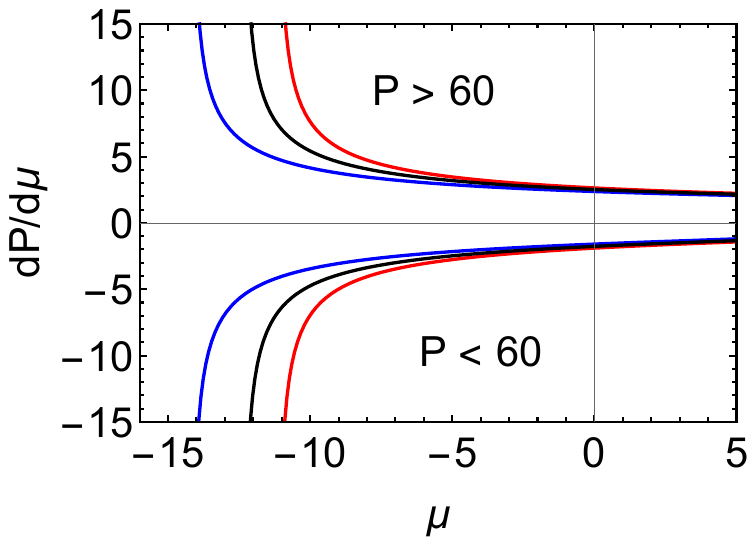} 
 	\caption{Left panel: Variation of effective potential with width for different values  of strength of optical lattice in presence of dipole-dipole interaction. The blue curve is drawn for $V_{0}=0$, black curve is drawn for $V_{0}=1$, and red curve is drawn $V_{0}=1.8$.  Right panel: {Variation of $\frac{dP}{d\mu}$ as a function of chemical potential $\mu$.}  In both the panel, we have taken $P=10,\,\, g_{\rm{e}}=4.75,\,\, g_{1}=0.75,\,\, C_{0}=1.7,\,\, k=1.5,\,\,\,x_{0}=0.5$ and $\eta=1.0$.   } 
 	\label{fig4}
 \end{figure}  
 \begin{figure}
 	\centering 
 	\includegraphics[scale=0.3]{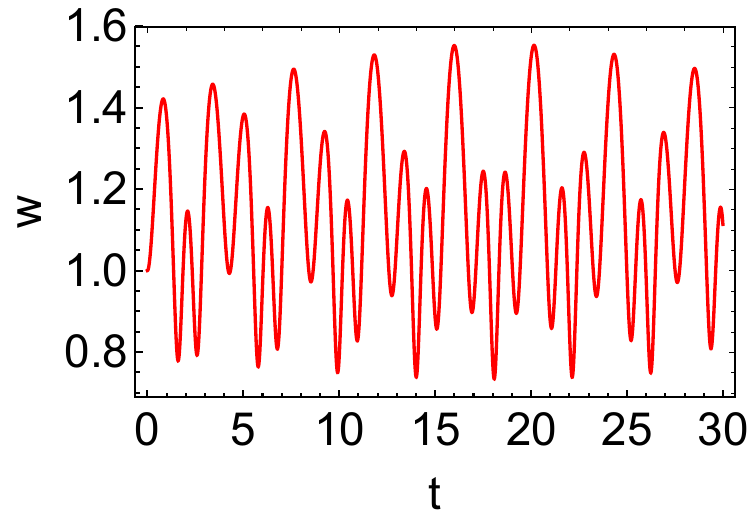}
 	\includegraphics[scale=0.3]{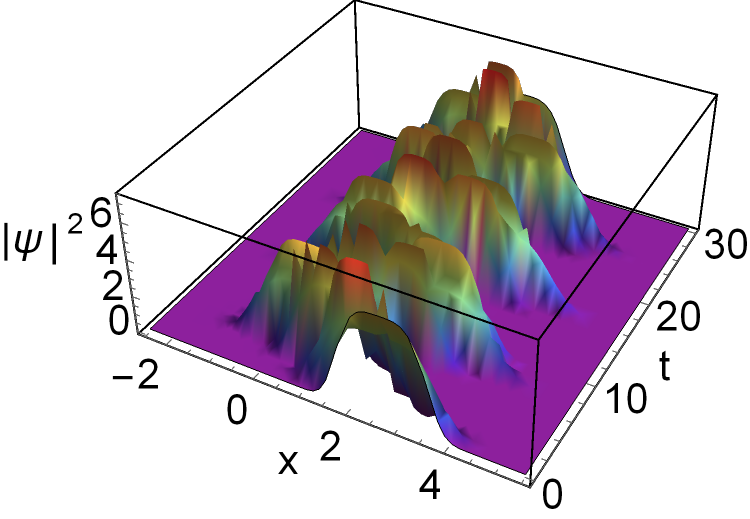}
 	\caption{ Left panel: Variation of  width with time in presence of both optical lattice and dipole-dipole interaction. Right panel: Evolution of density profile with time  for $x(0)=2.65$ and $w(0)=1.0$. In both the panel, we have taken $P=10,\,\, g_{\rm{e}}=4.75,\,\, g_{1}=0.75,\,\, C_{0}=1.7,\,\,\,V_{0}=1.8,\,\,\, k=1.5$ and $\eta=1.0$. } 
 	\label{fig5} 
 \end{figure}  
\begin{figure}
\centering 
\includegraphics[scale=0.325]{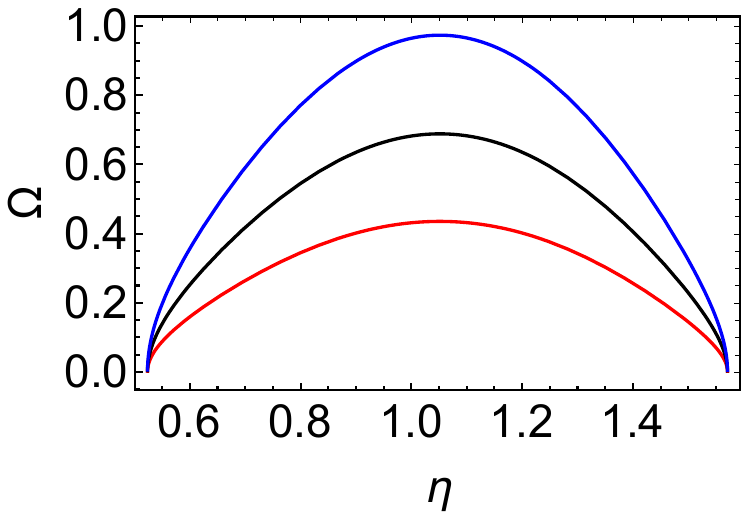}
\includegraphics[scale=0.325]{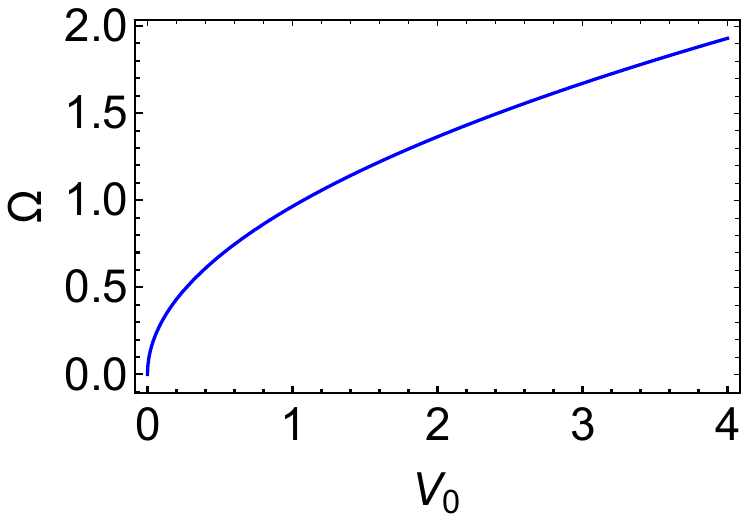}
\caption{Left panel: Frequency of oscillation ($\Omega$) of center of mass of a droplet with $\eta$ keeping $V_0$ fixed at $1.0$(blue curve), $0.5$ (black curve) and $0.2$ (red curve). Right panel: It gives $\Omega$ versus $V_0$ curve for $\eta=1.0$.  In both the panels, we take  $P=10,\,\, g_{\rm{e}}=4.75,\,\, g_{1}=0.75,$ and $ C_{0}=1.7$.} 
\label{fig6} 
\end{figure} 

With a view to study dynamical behavior of QDs in OL, we solve the coupled Eqs. (15) and (19)  for non-zero values of $V_0$ and $x_0$, and display the results in Fig. 5. 
In presence of both DDI and optical lattice, variation of width with time get modified. Particularly, the oscillation of width is not perfectly smooth. Indeed the motion is quasi-periodic (left panel). The density profile of a QD, however,  exhibits period motion in the lattice about its initial location (right panel). Here, the OL  accelerates center-of-mass of the droplet and increases frequency of oscillation of the density profile (Fig. 6).  Particularly, the growth of oscillation frequency with the increase of wave number is saturated at a particular value of wavenumber.  With the increase of lattice strength, the frequency of oscillation shows parabolic growth.

{The QDs in dipolar BECs with optical lattices are linearly stable. However, the variational approach  cannot accurately incorporate shape change of density profile during time evolution and thus unable to demand the dynamical stability. In order of this, we calculate  density profile with time through numerically solving the dimensionless GP equation in (7).  In particular, we consider split-step Fourier method  \cite{ssf1} for temporal step $\delta t\approx 0.001$ and spatial step $\delta x\approx 0.05$ with initial solution $\psi(x,0)=A(0) {\rm exp}[{-\frac{1}{2}{\left(\frac{x-x_0(0)}{w(0)}\right)}^{2\sigma}}]$.  In Fig. 7, we plot density profile (top panel) evolving near the center of the trap in absence of optical lattices. It shows that the amplitude and/or width of the profile is changing with time but profile is not spreading out. In the bottom panel, we compare the results of  width variation obtained from variational and numerical calculations and find qualitative agreement between the two results.}
\begin{figure}
\centering 
\includegraphics[scale=0.5]{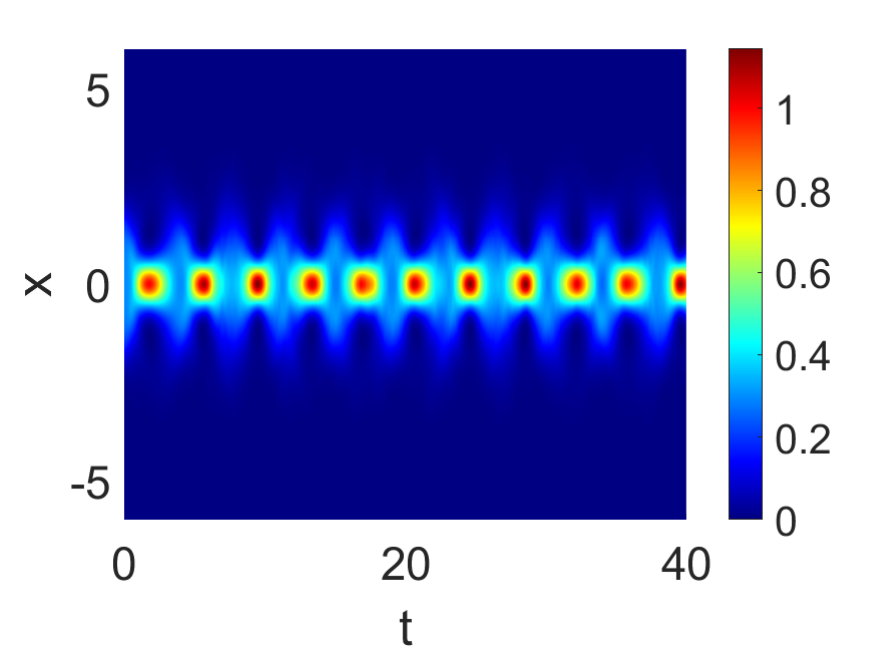}
\includegraphics[scale=0.45]{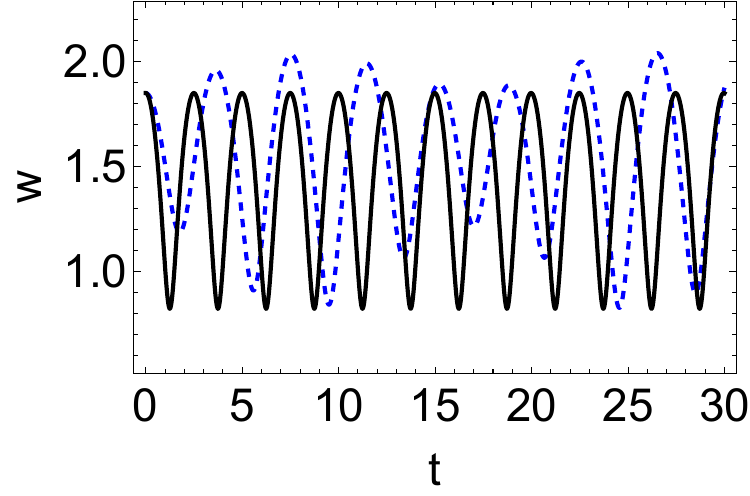}
\caption{Top panel: Evolution of density  profile of a QD near the center of the trap obtained from numerical calculation for  $V_0=0$, $P=10,\,\, g_{\rm{e}}=3.5,\,\, g_{1}=0.55,\,\, C_{0}=0.001$, $w(0)=1.85$ and $x_0(0)=0.005$. Bottom panel: Variation of width  with time. Here the dashed and solid curves give respectively numerical and variational results.} 
\label{fig7} 
\end{figure} 
{In presence of optical lattices, the evolution of density profile is affected and this effect crucially depends on the relative values of quantum fluctuation and mean-field interaction and also on the lattice environment.  In Fig. 8, we  display the time evolution of density profile in presence of optical lattices (top panel). This lattice causes the amplitude of density profile to vary relatively larger than that in absence of optical lattice. Interestingly, the density profile remains stable during evolution. However, its width is varying with time. We see that  the numerical results on the change of width   qualitatively match with  variational results.}
\begin{figure}
\centering 
\includegraphics[scale=0.5]{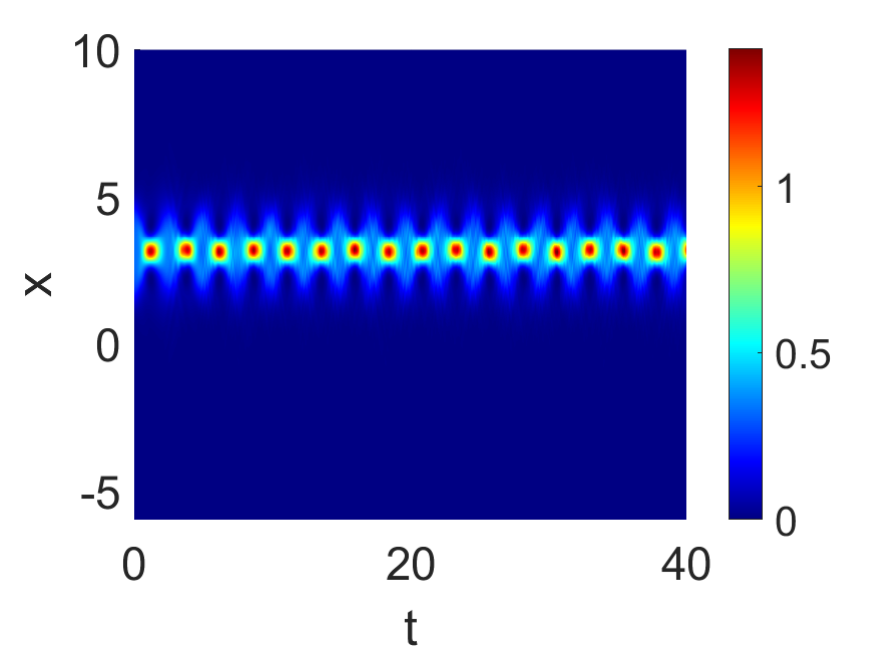}
\includegraphics[scale=0.45]{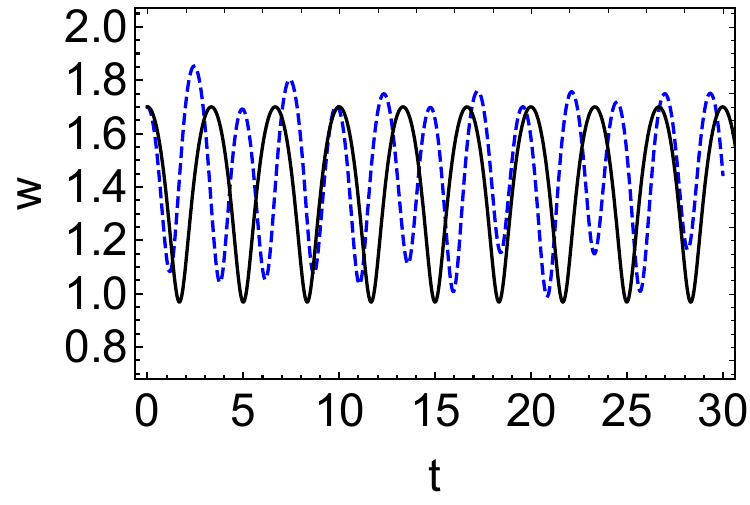}
\caption{Evolution of density  profile of a QD near the center of the trap obtained from numerical calculation for  $V_0=1.8$, $\eta=0.5$, $P=10,\,\, g_{\rm{e}}=4.75,\,\, g_{1}=0.75,\,\, C_{0}=0.0007$, $w(0)=1.7$ and $x_0(0)=3.1$. Bottom panel: Variation of width with time. Here the dashed and solid curves give respectively numerical and variational results.} 
\label{fig8} 
\end{figure} 

\section{Conclusion}  
In this work we have systematically studied stability and dynamics of quantum droplets (QDs) in dipolar quasi-one-dimensional Bose-Einstein condensates in presence of optical lattices. Based on variational analysis, we have found an effective potential for the width and shown that the optimum width (corresponding to the minimum of the potential) of QD increases with the increase of dipole-dipole interaction. The QD with optimum width satisfies the famous Vakhitov-Kolokolov criterion of linearly stability. The width of a stable QD oscillates with time and the amplitude of oscillation decreases with the increase of dipole-dipole interaction.  

In presence of optical lattices, the effective potential gets changed and results a change of optimal width of a QD.  We have shown that the QD can be linear stable for proper choices of lattice and interaction parameters. Interestingly, width of a stable QD oscillates with time quasi-periodically.   The frequency of oscillation depends sensitively on the wavenumber of the  lattice potential. However, its position oscillates periodically about the minimum of the lattice. We have examined through numerical simulation that the QDs can be stable during evolution both in presence and absence of optical lattices.

\subsection*{Acknowledgement:}  
	S. Siddik would like to thank "West Bengal Higher Education Department" for providing Swami Vivekananda Merit Cum Means Scholarship with F. No. WBP231685702213.


\begin{thebibliography}{99}  
		\bibitem{a1} D. S. Petrov, Quantum Mechanical Stabilization of a Collapsing Bose-Bose Mixture,  Phys. Rev. Lett. \textbf{115}, 155302 (2015). 
		\bibitem{a2}  T. D. Lee, K. Huang and C. N. Yang, Eigenvalues and Eigen functions of a Bose System of Hard Spheres	and Its Low-Temperature Properties,  Phys. Rev. \textbf{106}, 1135 (1957).  
		\bibitem{a6} C. R. Cabrera, L. Tanzi, J. Sanz, B. Naylor, P. Thomas,
		P. Cheiney, and L. Tarruell, Quantum liquid droplets in a mixture of Bose-Einstein condensate,  Science \textbf{359}, 301 (2018). 
		\bibitem{a7} G. Semeghini, G. Ferioli, L. Masi, C. Mazzinghi, L. Wolswijk, F. Minardi, M. Modugno, G. Modugno, M. Inguscio, and M. Fattori, Self-Bound Quantum Droplets of Atomic Mixtures in Free Space, Phys. Rev. Lett. \textbf{120}, 235301 (2018). 
		\bibitem{a8} C. D’Errico, A. Burchianti, M. Prevedelli, L. Salasnich,
		F. Ancilotto, M. Modugno, F. Minardi, and C. Fort, Observation of quantum droplets in a heteronuclear bosonic mixture,  Phys. Rev. Res. \textbf{1}, 033155 (2019).
		\bibitem{a9}  Z. Guo, F. Jia, L. Li, Y. Ma, J. M. Hutson, X. Cui, and D. Wang, Lee-Huang-Yang effects in the ultracold mixture of $^{23}Na$ and $^{87}Rb$ with attractive interspecies interactions,  Phys. Rev. Res. \textbf{3}, 033247 (2021).  
		\bibitem{a3} I. Ferrier-Barbut, H. Kadau, M. Schmitt, M. Wenzel, and T. Pfau, Observation of Quantum Droplets in a Strongly Dipolar Bose Gas,  Phys. Rev. Lett. \textbf{116}, 215301 (2016).
		\bibitem{pra1} F. W\"achtler and L. Santos, Ground-state properties and elementary excitations of quantum droplets in dipolar Bose-Einstein condensates Phys. Rev. A {\bf 94}, 043618(2016).
		\bibitem{pra2} F. W\"achtler and L. Santos, Quantum filaments in dipolar Bose-Einstein condensates, Phys. Rev. A {\bf 94}, 021602(R)(2016).
		\bibitem{pra3} F. B{\"o}ttcher, M. Wenzel, Jan-Niklas Schmidt, M. Guo, T. Langen, I. Ferrier-Barbut, and T. Pfau, R. Bombin, J. Sanchez-Baena, J. Boronat, and F. Mazzanti, Dilute dipolar quantum droplets beyond the extended Gross-Pitaevskii equation, Phys. Rev. Research {\bf 1}, 033088(2019).
		\bibitem{a4}  M. Schmitt, M. Wenzel, F. B¨ottcher, I. Ferrier-Barbut, and T. Pfau, Self-bound droplets of a dilute magnetic quantum liquid,   Nature \textbf{539}, 259 (2016). 
		\bibitem{a5} L. Chomaz, S. Baier, D. Petter, M. J. Mark, F. W\"achtler,
		L. Santos, and F. Ferlaino, Quantum-fluctuation-driven crossover from a dilute Bose-Einstein condensate to a macro-droplet in a dipolar quantum fluid,  Phys. Rev. X \textbf{6}, 041039 (2016). 
		\bibitem{nr1} A. Cidrim, F. E. A. dos Santos, E. A. L. Henn, and T. Macri, Vortices in self-bound dipolar droplets, Phys. Rev. A \textbf{98}, 023618 (2018).
		
\bibitem{nr2} H. Huang, H. Wang, G. Chen, M. Chen, C. S. Lim, and K.-C. Wong, Stable quantum droplets with higher-order vortex in radial lattices, Chaos Solit. Fractals \textbf{168}, 113137 (2023). 

\bibitem{nr3} G. Li, Z. Zhao, X. Jiang, Z. Chen, B. Liu, B.A. Malomed and Y. Li,
Strongly anisotropic vortices in dipolar quantum droplets Phys. Rev. Lett. \textbf{133}, 053804 (2024).
\bibitem{nr4} F. Bottcher, J.-N. Schmidt, J. Hertkorn, K.S. Ng, S.D. Graham, M. Guo, T. Langen, T. Pfau, New states of matter with fine-tuned interactions: quantum droplets and dipolar supersolids
Rep. Progr. Phys. \textbf{84}, 012403(2020).
\bibitem{nr5} Fabian Bottcher, Jan-Niklas Schmidt, Matthias Wenzel, Jens Hertkorn, Mingyang Guo,Tim Langen, and Tilman Pfau, Transient supersolid properties in an array of dipolar quantum droplets, Phys. Rev. X \textbf{9}, 011051 (2019).
\bibitem{nr6} L. Tanzi, E. Lucioni, F. Fama, J. Catani,  A. Fioretti, C. Gabbanini, R. N. Bisset, L. Santos, and G. Modugno,  Observation of a dipolar quantum gas with metastable supersolid properties, Phys. Rev. Lett. \textbf{122}, 130405(2019).
\bibitem{bb17} M. Greiner, S. Folling, Condensed-matter physics,  Optical lattice. Nature \textbf{453}, 736 (2009).  
\bibitem{aa11} K. Goral, L. Santos, and M. Lewenstein, Quantum phases of dipolar bosons in optical lattices, Phys. Rev. Lett. \textbf{88}, 170406 (2002).
\bibitem{aa12} H. P. Buchler, E. Demler, M. Lukin, A. Micheli, N. Prokof'ev, G. Pupillo, and P. zoller, Strongly correlated 2D quantum phases with cold polar molecules: controlling the shape of the interaction potential, Phys. Rev. Lett. \textbf{98}, 060404 (2007).
\bibitem{nr7} R. Kraus, T. Chanda, J. Zakrzewski, and G. Morigi, Quantum phases of dipolar bosons in one-dimensional optical lattices, Phys. Rev. B \textbf{106}, 035144 (2022).
\bibitem{nr8}  Zheng Zhou, Xi Yu, Yu Zou, Honghua Zhong, Dynamics of quantum droplets in a one-dimensional optical lattice,  Commun Nonlinear Sci Numer Simulat \textbf{78}, 104881(2019).
\bibitem{nr9} L. Dong,  W. Qi, P. Peng, L. Wang, H. Zhou and C. Huang, Multi-stable quantum droplets in optical lattices, Nonlinear Dyn. \textbf{102}, 310(2020).
\bibitem{nr9a}  Yuhang Nie, Jun-Hui Zheng and Tao Yang, Spectra and dynamics of quantum droplets in an optical lattice, Phys. Rev. A \textbf{108},053310(2023.
\bibitem{nr10} M. G. Vakhitov, A. A. Kolokolov, Stationary solutions of the wave equation in a medium with nonlinearity saturation. Radiophys Quantum Electron \textbf{16}, 783(1973). 
\bibitem{nr11} Sk. Golam Ali, S. K. Roy and B. Talukdar, Stability of matter-wave solitons in optical lattices, Eur. Phys. J. D \textbf{59}, 269(2010).
\bibitem{nr12} Sudipta Das and Golam Ali Sekh, Dynamics of compressed optical pulse in cubic-quintic media, Fiber \& Integrated  Optics \textbf{39}, 122(2020) 
\bibitem{f1} D. S. Petrov, G. V. Shlyapnikov, and J. T. M. Walraven,Regimes of quantum degeneracy in trapped 1D gases, Phys. Rev. Lett.\textbf{85}, 3745 (2000).
\bibitem{f2}  S. R. Otajonov, B. A. Umarov and F. Kh. Abdullaev, Modulational instability and discrete quantum droplets in a deep quasi-one-dimensional optical lattice, Phys. Rev. E \textbf{111}, 054206(2025).
\bibitem{f3} S. R. Otajonov, B. A. Umarov, and F. K. Abdullaev, Dynamics of quasi-one-dimensional quantum droplets in Bose-Bose mixtures,  Chaos, Solitons and Fractals \textbf{186}, 115212 (2024).
\bibitem{f4}  F. Kh. Abdullaev, R. M. Galimzyanov, A. M. Shermakhmatov, Effects of quantum fluctuations on macroscopic quantum tunnelling and self-trapping of BEC in a double well trap. J. Phys. B: At. Mol. Opt. Phys. \textbf{56}, 165301 (2023).
\bibitem{f4a} Matthew Edmonds, Thomas Bland and Nick Parker, Quantum droplets of quasi-one-dimensional dipolar Bose–Einstein condensates, J. Phys. Commun. \textbf{4}, 125008 (2020).
{\bibitem{f34} Argha Debnath and Ayan Khan, Investigation of quantum Droplets: An analytical approach, Ann. Phys.  (Berlin) \textbf{533}, 2000549 (2021).}
\bibitem{f4b} Ralf Schutzhold, Michael Uhlmann, Yan Xu, And Uwe R. Fischer, Mean-field expansion in Bose–Einstein condensates with finite-range interactions, Int. J.  Mod.  Phys.  B   \textbf{20}, 3565(2006).
\bibitem{f5}  F. Deuretzbacher, J. C. Cremon, and S. M. Reimann, Ground-state properties of few dipolar bosons in a quasi-one-dimensional harmonic trap, Phys. Rev. A. \textbf{81}, 063616 (2010).
\bibitem{f6} Y. Cai, M. Rosenkranz, Z. Lei and W. Bao, Mean-field regime of trapped dipolar Bose-Einstein condensates in one and two dimensions,  Phys. Rev. A \textbf{82}, 043623 (2010). 
\bibitem{f7} M. Edmonds, T. Bland and N. Parker, Stability of intrinsic localized modes on the lattice with competing power non-linearity,  J. Phys. Commun. \textbf{4}, 125008 (2020). 
{\bibitem{36a} S. Sinha and L. Santos, Cold dipolar gases in Q1D geometrics, Phys. Rev. Lett. \textbf{99}, 140406 (2007).} 
{\bibitem{36b} L. Khaykovich, F. Schreck, G. Ferrari,  T. Bourdel, J. Cubizolles, L. D. Carr, Y. Castin,C. Salomon, Formation of a matter-wave bright soliton,  Science \textbf{296}, 1290(2002).} 
\bibitem{f8} S. R. Otajonov, Dynamics of quasi-one-dimensional quantum droplets in Bose-Bose mixtures, J. Phys. B: At. Mol. Opt. Phys. \textbf{55}, 085001 (2022).
\bibitem{nr13}  Sk Siddik and Golam Ali Sekh, Dynamics of quantum droplets in Bose-Einstein condensates with three-body loss and external feeding,  Commun. Theor. Phys. \textbf{78}, 055501(2026).
{\bibitem{ssf1} X. Antoine and R. Duboscq, GPELab, a Matlab Toolbox to Solve Gross-Pitaevskii Equations II: Dynamics and Stochastic Simulations, Comp. Phys. Commun. \textbf{193}, 95(2015).}
\end{thebibliography}
\end{document}